\documentclass{article}

\usepackage{amsmath,graphicx,subfig}
\usepackage{amssymb,amsfonts}
\usepackage{overpic}
\usepackage[cal=boondox]{mathalpha}
\usepackage{bm}
\usepackage{enumitem}
\graphicspath{ {./images/} }
\usepackage{multirow}
\usepackage{array}
\usepackage{mdwmath}
\usepackage{mdwtab}
\usepackage{booktabs}
\usepackage[nodisplayskipstretch]{setspace} 
\usepackage[ruled,vlined,lined,commentsnumbered]{algorithm2e}
\usepackage[compact]{titlesec}
\usepackage{arxiv}

\usepackage[utf8]{inputenc} % allow utf-8 input
\usepackage[T1]{fontenc}    % use 8-bit T1 fonts
\usepackage{hyperref}       % hyperlinks
\usepackage{url}            % simple URL typesetting
\usepackage{booktabs}       % professional-quality tables
\usepackage{amsfonts}       % blackboard math symbols
\usepackage{nicefrac}       % compact symbols for 1/2, etc.
\usepackage{microtype}      % microtypography
\usepackage{lipsum}		% Can be removed after putting your text content
\usepackage{graphicx}
\usepackage{doi}
\usepackage{csquotes}

\usepackage{tikz}
\usepackage{textcomp}

\title{Complex-Valued Neural Networks for Data-Driven Signal Processing and Signal Understanding}

\author{
        \href{https://orcid.org/0000-0002-3388-4805}{\includegraphics[scale=0.06]{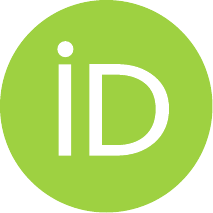}\hspace{1mm}Josiah W. Smith} \\
	\texttt{josiah.radar@gmail.com} \\
}

\date{}

\renewcommand{\headeright}{Preprint}
\renewcommand{\undertitle}{Preprint}
\renewcommand{\shorttitle}{CVNNs for Data-Driven Signal Processing and Signal Understanding}

\hypersetup{
pdftitle={Complex-Valued Deep Learning for Signal Processing},
pdfsubject={cs.AI, eess.SP},
pdfauthor={Josiah W.~Smith},
pdfkeywords={Complex-Valued Neural Networks (CVNNs), Deep Learning, Artificial Intelligence, Signal Processing, Radar, Synthetic Aperture Radar},
}

\begin{document}
\maketitle

\begin{abstract}
Complex-valued neural networks have emerged boasting superior modeling performance for many tasks across the signal processing, sensing, and communications arenas. 
However, developing complex-valued models currently demands development of basic deep learning operations, such as linear or convolution layers, as modern deep learning frameworks like PyTorch and Tensor flow do not adequately support complex-valued neural networks. 
This paper overviews a package built on \href{https://pytorch.org/}{PyTorch} with the intention of implementing light-weight interfaces for common complex-valued neural network operations and architectures. 
Similar to natural language understanding (NLU), which as recently made tremendous leaps towards text-based intelligence, \textit{RF Signal Understanding (RFSU)} is a promising field extending conventional signal processing algorithms using a hybrid approach of signal mechanics-based insight with data-driven modeling power. 
Notably, we include efficient implementations for linear, convolution, and attention modules in addition to activation functions and normalization layers such as batchnorm and layernorm.
Additionally, we include efficient implementations of manifold-based complex-valued neural network layers that have shown tremendous promise but remain relatively unexplored in many research contexts. 
Although there is an emphasis on 1-D data tensors, due to a focus on signal processing, communications, and radar data, many of the routines are implemented for 2-D and 3-D data as well.
Specifically, the proposed approach offers a useful set of tools and documentation for data-driven signal processing research and practical implementation. 
\end{abstract}

% keywords can be removed

\section{Introduction}
\label{sec:intro}

Although much work as already been done for complex-valued neural networks (CVNNs), starting back in the 1990s \cite{kim1990modification}, many common operations for complex-valued tensors remain unsupported by modern deep learning frameworks like PyTorch and TensorFlow. 
In this paper, we introduce a lightweight wrapper built on PyTorch with two main objectives as follows
\begin{itemize}
    \item Provide an efficient interface for complex-valued deep learning using PyTorch.
    \item Provide open-source implementations and documentation for complex-valued operation such activation functions, normalization layers, and attention modules to increase the research rate for CVNNs.
\end{itemize}

Fundamentally, complex-valued data have two degrees of freedom (DOF), which are commonly modeled as real and imaginary parts or magnitude and phase as
\begin{equation}
    \label{eq:basic_complex}
    \mathbf{z} = \mathbf{x} + j\mathbf{y} = |\mathbf{z}| e^{j \angle \mathbf{z}},
\end{equation}
where $j = \sqrt{-1}$ is known as the complex unit, $\mathbf{x}$ and $\mathbf{y}$ are the real and imaginary parts of $\mathbf{z}$, and $|\mathbf{z}|$ and $\angle \mathbf{z}$ are the magnitude and phase of $\mathbf{z}$. 

\subsection{Existing CVNN Work}
\label{subsec:background_literature}

Recent surveys of CVNNs and their history can be found in \cite{lee2022complex,bassey2021survey}. 
Starting in 2018, Trabelsi \textit{et al.} introduced \textit{deep complex networks} (DCN) extending many common deep learning approaches to complex-valued data \cite{trabelsi2018deep}. 
As mentioned in \cite{barrachina2023theory}, some attempts have been made by PyTorch and TensorFlow to incorporate complex-values into the core architecture; however, much work remains to be done. 
The authors of \cite{barrachina2023theory} implement a CVNN framework in TensorFlow, with impressive functionality. 
However, recent studies have shown a drastic decline in TensorFlow usage among engineers, while PyTorch dominates industry and academia. 
The code publicized as part of this work was used extensively throughout the dissertation \cite{smith2022novel} in addition to \cite{smith2020nearfieldisar,smith2021sterile,smith2021An,smith2022efficient,smith2022ffh_vit,vasileiou2022efficient,smith2023dual_radar,smith2023ThzToolbox,smith2023dualradarsarcontroller}. 

One paramount issue for CVNNs is complex-valued backpropagation and computation of complex-valued gradients. 
Mathematical investigations for Liouville Theorem and Wirtinger Calculus can be found in \cite{barrachina2023theory,trabelsi2018deep}, and are omitted here for brevity. 
Whereas, \cite{barrachina2023theory} develops Wirtinger-based proper backpropagation for TensorFlow, PyTorch natively supports complex-valued backpropagation. 

\subsection{Installation Notes}
\textbf{IMPORTANT:} Prior to installation, \href{https://pytorch.org/get-started/locally/}{install PyTorch} to your environment using your preferred method using the compute platform (CPU/GPU) settings for your machine. PyTorch will not be automatically installed with the installation of \texttt{complextorch} and MUST be installed manually by the user.

\subsubsection{Install using PyPI}
\texttt{complextorch} is available on the Python Package Index (PyPI) and can be installed using the following command:

\texttt{pip install complextorch}

\subsubsection{Install using GitHub}
Useful if you want to modify the source code.

\texttt{git clone https://github.com/josiahwsmith10/complextorch.git}

\section{Complex-Valued Layers}
\label{subsec:complex_valued_layers}

In this section, we overview the complex-valued layers supported by \texttt{complextorch}, which can be found at \href{https://complextorch.readthedocs.io/en/latest/nn/modules.html}{https://complextorch.readthedocs.io/en/latest/nn/modules.html}. 
Nearly all modules introduced in this section closely follow the PyTorch to improve integration and streamline user experience (UX) for new users.

\subsection{Gauss' Multiplication Trick}
\label{subsec:gauss_multiplication_trick}

As pointed out in \cite{chakraborty2019sur_real,chakraborty2020C_SURE,chakraborty2022SurReal}, many complex-valued operations can be more efficiently implemented by leveraging Gauss' multiplication trick. 
Suppose $\mathcal{L}(\cdot) = \mathcal{L}_\mathbb{R}(\cdot) + j \mathcal{L}_\mathbb{I}(\cdot)$ is a linear operator, such as multiplication or convolution, and $\mathbf{z} = \mathbf{x} + j\mathbf{y}$. 
Hence,
\begin{equation}
    \label{eq:gauss1}
    \mathcal{L}(\mathbf{z}) = \mathcal{L}(\mathbf{x}) + j\mathcal{L}(\mathbf{y}) = \mathcal{L}_\mathbb{R}(\mathbf{x}) - \mathcal{L}_\mathbb{I}(\mathbf{y}) + j \left(\mathcal{L}_\mathbb{R}(\mathbf{y}) + \mathcal{L}_\mathbb{I}(\mathbf{x})\right).
\end{equation}
This is the common implementation of complex-valued operations in deep learning applications. 
However, it requires four computations, such as multiplications or convolutions, which can become computationally costly. 
Gauss' trick reduces the number of computations down to 3 as
\begin{equation}
    \begin{gathered}
        t_1 \triangleq \mathcal{L}_\mathbb{R}(\mathbf{x}), \\
        t_2 \triangleq \mathcal{L}_\mathbb{I}(\mathbf{y}), \\
        t_3 \triangleq (\mathcal{L}_\mathbb{R} + \mathcal{L}_\mathbb{I})(\mathbf{x} + \mathbf{y}), \\
    \end{gathered}
\end{equation}
\begin{equation}
    \label{eq:gauss_trick}
    \mathcal{L}(\mathbf{z}) = t_1 - t_2 + j(t_3 - t_2 - t_1).
\end{equation}
This technique is leveraged throughout \texttt{complextorch} to improve computational efficiency whenever applicable. 

\subsection{Complex-Valued Linear Layers}
\label{subsec:cvlinear}

We extend linear layers, the backbone of perceptron neural networks, for CVNNs. 
Similar to Section \ref{subsec:gauss_multiplication_trick}, we define the complex-valued linear layer \texttt{CVLinear} as 
\begin{equation}
    \label{eq:cvlinear}
    H_\texttt{CVLinear}(\cdot) = H_{\texttt{CVLinear-}\mathbb{R}}(\cdot) + j H_{\texttt{CVLinear-}\mathbb{I}}(\cdot),
\end{equation}
where $H_{\texttt{CVLinear-}\mathbb{R}}(\cdot)$ and $H_{\texttt{CVLinear-}\mathbb{I}}(\cdot)$ can be implemented in PyTorch as real-valued \href{https://pytorch.org/docs/stable/generated/torch.nn.Linear.html}{Linear} layers, as detailed in Table \ref{tab:cvlinear}. 
Using \eqref{eq:gauss_trick}, the linear layer can be efficiently computed with the composition $H_{\texttt{CVLinear-}\mathbb{R}}(\cdot) + H_{\texttt{CVLinear-}\mathbb{I}}(\cdot)$ being a linear layer with the weights and bias of $H_{\texttt{CVLinear-}\mathbb{R}}(\cdot)$ and $H_{\texttt{CVLinear-}\mathbb{I}}(\cdot)$ summed.

\begin{table}[h]
    \centering
    \begin{tabular}{c|c}
        \href{https://complextorch.readthedocs.io/en/latest/index.html}{\texttt{complextorch}} & \href{https://pytorch.org/}{PyTorch}  \\
        \hline
        \texttt{CVLinear} & \texttt{Linear}
    \end{tabular}
    \caption{PyTorch equivalent of the complex-valued linear layer.}
    \label{tab:cvlinear}
\end{table}

\subsection{Complex-Valued Convolution Layers}
\label{subsec:cvconv}

Similarly, we define a general complex-valued convolution layer \texttt{CVConv} as 
\begin{equation}
    \label{eq:CVConv}
    H_\texttt{CVConv}(\cdot) = H_{\texttt{CVConv-}\mathbb{R}}(\cdot) + j H_{\texttt{CVConv-}\mathbb{I}}(\cdot),
\end{equation}
where $H_{\texttt{CVConv-}\mathbb{R}}(\cdot)$ and $H_{\texttt{CVConv-}\mathbb{I}}(\cdot)$ can be implemented in PyTorch as various real-valued convolution layers, as detailed in Table \ref{tab:cvlinear}. 
Using \eqref{eq:gauss_trick}, the convolution layers layer can be efficiently computed with the composition $H_{\texttt{CVConv-}\mathbb{R}}(\cdot) + H_{\texttt{CVConv-}\mathbb{I}}(\cdot)$ being a convolution layer with the kernel weights and bias of $H_{\texttt{CVConv-}\mathbb{R}}(\cdot)$ and $H_{\texttt{CVConv-}\mathbb{I}}(\cdot)$ summed.

\begin{table}[h]
    \centering
    \begin{tabular}{c|c}
        \href{https://complextorch.readthedocs.io/en/latest/index.html}{\texttt{complextorch}} & \href{https://pytorch.org/}{PyTorch}  \\
        \hline
        \texttt{CVConv1d} & \texttt{Conv1d} \\
        \texttt{CVConv2d} & \texttt{Conv2d} \\
        \texttt{CVConv3d} & \texttt{Conv3d} \\
        \texttt{CVConvTranpose1d} & \texttt{ConvTranpose1d} \\
        \texttt{CVConvTranpose2d} & \texttt{ConvTranpose2d} \\
        \texttt{CVConvTranpose3d} & \texttt{ConvTranpose3d} \\
    \end{tabular}
    \caption{PyTorch equivalent of complex-valued convolution layers.}
    \label{tab:cvconv}
\end{table}

\subsection{Complex-Valued Attention Layers}
\label{subsec:cvattention}
Whereas attention-based models, such as transformers, have gained significant attention for natural language processing (NLP) and image processing, their potential for implementation in complex-valued problems such as signal processing remains relatively untapped. 
Here, we include complex-valued variants of several attention-based techniques.

\subsubsection{Complex-Valued Scaled Dot-Product Attention}
\label{subsubsec:cvattention_mechanism}
The ever-popular scaled dot-product attention is the backbone of many attention-based methods \cite{vaswani2017attention}, most notably the transformer \cite{dosovitskiy2020image_ViT,mehta2021mobilevit,liu2021swin,liu2021swinv2,liang2021swinir,smith2022ffh_vit}.

Given complex-valued \textit{query, key, and value} tensors $Q, K, V$, the complex-valued scaled dot-product attention can be computed as
\begin{equation}
    \label{eq:complex_attention_mechanism}
    \text{Attention}(Q,K,V) = \mathcal{S}(QK^T/t)V,
\end{equation}
where $t$ is known as the temperature (typically $t = \sqrt{d_{attn}}$) and $\mathcal{S}$ is the softmax function. 

It is important to note that unlike real-valued scaled dot-product attention, the complex-valued version detailed above must employ a complex-valued version of the soft-max function as the real-valued softmax is unsuited for complex-valued data. 
We implemented several complex-valued softmax function options detailed in Section \ref{subsec:cvsoftmax}. 

Additionally, we include an implementation of multi-head attention, which is commonly employed in transformer and attention-based models \cite{liu2021swin}. 

\subsubsection{Complex-Valued Efficient Channel Attention (CV-ECA)}
\label{subsubsec:cveca}

Efficient Channel Attention (ECA) was first introduced in \cite{wang2020AnEfficientCSA}. 
Here, we extend ECA to complex-valued data for CV-ECA. 
Following the construction of ECA, we define CV-ECA as
\begin{equation}
    \label{eq:cveca}
    \texttt{CV-ECA} = \mathcal{M}\left( H_\texttt{CVConv1d}(H_\texttt{CVAdaptiveAvgPoolNd}(\mathbf{z})) \right) \odot \mathbf{z},
\end{equation}
where $\mathcal{M}(\cdot)$ is the masking function (some options are implemented in Sections \ref{subsec:cvsoftmax} and \ref{subsec:cvmask}), $H_\texttt{CVConvNd}(\cdot)$ is the 1-D complex-valued convolution layer defined in Section \ref{subsec:cvconv}, and $H_\texttt{CVAdaptiveAvgPoolNd}(\cdot)$ is the complex-valued global adaptive average pooling layer for the $N$-D input tensor detailed in Section \ref{subsec:cvpool}. 
It is important to note that the 1-D convolution is computed along the channel dimension of pooled data. 
This is the notable difference between ECA and MCA. 

\subsubsection{Complex-Valued Masked Channel Attention (CV-MCA)}
\label{subsubsec:cvmasked_attention}

Complex-valued masked channel attention (CV-MCA) is similar to CV-ECA but employs a slightly different implementation. 
Generally, the masked attention module implements the following operation
\begin{equation}
    \label{eq:masked_attention}
    \texttt{CV-MCA}(\mathbf{z}) = \mathcal{M}(H_\text{ConvUp}(\mathcal{A}(H_\text{ConvDown}(\mathbf{z})))) \odot \mathbf{z},
\end{equation}
where $\mathcal{M}(\cdot)$ is the masking function (some options are implemented in Sections \ref{subsec:cvsoftmax} and \ref{subsec:cvmask}), $H_\text{ConvUp}(\cdot)$ and $H_\text{ConvDown}(\cdot)$ are $N$-D convolution layers with kernel sizes of 1 that reduce the channel dimension by a factor $r$, and $\mathcal{A}(\cdot)$ is the complex-valued non-linear activation layer (see Section \ref{subsec:cvact}). 
The implementations of complex-valued masked attention are included for 1-D, 2-D, and 3-D data. 
For more information on complex-valued masked channel attention, see the paper that introduced it \cite{cho2021complex}. 

\subsection{Complex-Valued Softmax Layers}
\label{subsec:cvsoftmax}

Softmax is an essential function for several tasks throughout deep learning. 
Complex-valued softmax functions have not been explored thoroughly in the literature at the time of this paper. 
However, a similar route, know as masking, has seen some attention in recent research \cite{cho2021complex}. 

Here, we introduce several softmax layer suitable for complex-valued tensors. 

\subsubsection{Split Type-A Complex-Valued Softmax Layer}
Using the definition of a split Type-A function from Section \ref{subsubsec:typeA}, we define the complex-valued split Type-A softmax layer (\texttt{CVSoftMax}) as
\begin{equation}
    \label{eq:cvsoftmax}
    \texttt{CVSoftMax}(\mathbf{z}) = \texttt{SoftMax}(\mathbf{x}) + j \texttt{SoftMax}(\mathbf{y}),
\end{equation}
where $\mathbf{z} = \mathbf{x} + j\mathbf{y}$ and \texttt{SoftMax} is the PyTorch real-valued softmax function. 

\subsubsection{Phase Preserving Softmax}
Similar to a Type-B function from Section \ref{subsubsec:typeB}, we define the complex-valued phase preserving softmax layer (\texttt{PhaseSoftmax}) as
\begin{equation}
    \label{eq:phasesoftmax}
    \texttt{CVSoftMax}(\mathbf{z}) = \texttt{SoftMax}(|\mathbf{z}|) \odot \frac{\mathbf{z}}{|\mathbf{z}|},
\end{equation}
where $\mathbf{z} = \mathbf{x} + j\mathbf{y}$ and \texttt{SoftMax} is the PyTorch real-valued softmax function.

\subsubsection{Magnitude Softmax}
We define the magnitude softmax layer, which simply computes the softmax over the magnitude of the input and ignores the phase, (\texttt{MagSoftmax}) as
\begin{equation}
    \label{eq:magsoftmax}
    \texttt{CVSoftMax}(\mathbf{z}) = \texttt{SoftMax}(|\mathbf{z}|),
\end{equation}
where $\mathbf{z} = \mathbf{x} + j\mathbf{y}$ and \texttt{SoftMax} is the PyTorch real-valued softmax function. 

\subsection{Complex-Valued Masking Layers}
\label{subsec:cvmask}

\subsubsection{Complex Ratio Mask (cRM) or Phase Preserving Sigmoid}
Detailed in Eq. (23) of \cite{cho2021complex}, the complex ratio mask (cRM) applies the traditional sigmoid function to the magnitude of the signal while leaving the phase information unchanged as
\begin{equation}
    \label{eq:complexratiomask}
    \texttt{ComplexRatioMask}(\mathbf{z}) = \texttt{Sigmoid}(|\mathbf{z}|) \odot \frac{\mathbf{z}}{|\mathbf{z}|}
\end{equation}
where $\mathbf{z} = \mathbf{x} + j\mathbf{y}$ and \texttt{Sigmoid} is the PyTorch real-valued softmax function. 

\subsubsection{Magnitude Min-Max Normalization Layer}
The min-max norm, which has proven useful in complex-value data normalization \cite{smith2023dual_radar}, is another option for a complex-valued softmax function as
\begin{equation}
    \label{eq:minmaxnorm}
    \texttt{MagMinMaxNorm}(\mathbf{z}) = \frac{\mathbf{z} - \mathbf{z}_\text{min}}{\mathbf{z}_\text{max} - \mathbf{z}_\text{min}}
\end{equation}
where $\mathbf{z} = \mathbf{x} + j\mathbf{y}$ and $\mathbf{z}_\text{min}$ and $\mathbf{z}_\text{max}$ indicate the minimum and maximum of $|\mathbf{z}|$. 

\subsection{Complex-Valued Normalization Layers}
\label{subsec:cvnorm}

Normalization layers are a crucial aspect of modern deep learning algorithms facilitating improved convergence and model robustness. 
As discussed in \cite{trabelsi2018deep}, traditional standardization of complex-valued data is not sufficient to translate and scale the data to unit variance and zero mean. 
Rather, a whitening procedure is necessary to ensure a circular distribution with equal variance for the real and imaginary parts of the signal. 
The whitening algorithm is derived in greater detail in \cite{trabelsi2018deep}, but we employ the same procedure for both batch normalization and layer normalization. 

\subsection{Complex-Valued Activation Layers}
\label{subsec:cvact}

Complex-valued activation functions are a key element of CVNNs that different significantly from real-valued neural networks. 
The popular real-valued activation layers (such as ReLU and GeLU) cannot be directly applied to complex-valued data without some modification. 
Complex-valued activation functions must take into account the 2 degrees-of-freedom inherent to complex-valued data, typically represented as real and imaginary parts or magnitude and phase.

We highlight four categories of complex-valued activation layers:
\begin{itemize}
    \item Split Type-A Activation Layers
    \item Split Type-B Activation Layers
    \item Fully Complex Activation Layers
    \item ReLU-Based Complex-Valued Activation Layers
\end{itemize}

Split \textit{Type-A} and \textit{Type-B} activation layers apply real-valued activation functions to either the real and imaginary or magnitude and phase, respectively, of the input signal \cite{lee2022complex,barrachina2023theory}. 
Fully complex activation layers are entirely complex-valued, while ReLU-based complex-valued activation layers are the family of complex-valued activation functions that extend the ever-popular ReLU to the complex plane. 

\subsubsection{Split Type-A Complex-Valued Activation Layers}
\label{subsubsec:typeA}

\textit{Type-A} activation functions consist of two real-valued functions, $G_\mathbb{R}(\cdot)$ and $G_\mathbb{I}(\cdot)$, which are applied to the real and imaginary parts of the input tensor, respectively, as 
\begin{equation}
    \label{eq:typeA}
    G(\mathbf{z}) = G_\mathbb{R}(\mathbf{x}) + j G_\mathbb{I}(\mathbf{y}).
\end{equation}
In most cases, $G_\mathbb{R}(\cdot) = G_\mathbb{I}(\cdot)$; however, $G_\mathbb{R}(\cdot)$ and $G_\mathbb{I}(\cdot)$ can also be distinct functions. 
Table \ref{tab:typeA} details the Type-A activation functions included in \texttt{complextorch}. 
Additionally, a generalized Type-A activation function is included allowing the user to adopt any set of $G_\mathbb{R}(\cdot)$ and $G_\mathbb{I}(\cdot)$ desired. 

\begin{table}[ht]
    \centering
    \begin{tabular}{c|c|c}
        \href{https://complextorch.readthedocs.io/en/latest/index.html}{\texttt{complextorch}} Activation Layer & $G_\mathbb{R}(\mathbf{z}) = G_\mathbb{I}(\mathbf{z})$ & Reference \\
        \hline
        \texttt{CVSplitTanh} & $\tanh(\mathbf{z})$ & Eq. (15) \cite{hirose2012generalization} \\
        \hline
        \texttt{CTanh} & $\tanh(\mathbf{z})$ & Eq. (15) \cite{hirose2012generalization} \\
        \hline
        \texttt{CVSplitSigmoid} & $\sigma(\mathbf{z})$ & - \\
        \hline
        \texttt{CSigmoid} & $\sigma(\mathbf{z})$ & - \\
        \hline
        \texttt{CVSplitAbs} & $|\mathbf{z}|$ & Section III-C \cite{marseet2017application} \\
        \hline
    \end{tabular}
    \caption{Type-A activation functions.}
    \label{tab:typeA}
\end{table}

\subsubsection{Polar Type-B Complex-Valued Activation Layers}
\label{subsubsec:typeB}

Similarly, \textit{Type-B} activation functions consist of two real-valued functions, $G_{||}(\cdot)$ and $G_\angle(\cdot)$, which are applied to the magnitude (modulus) and phase (angle, argument) of the input tensor, respectively, as
\begin{equation}
    \label{eq:typeB}
    G(\mathbf{z}) = G_{||}(|\mathbf{z}|) \odot \exp(j G_\angle(\angle\mathbf{z})).
\end{equation} 
Table \ref{tab:typeB} details the Type-B activation functions included in \texttt{complextorch}. 
Where $G_\angle(\cdot)$ is omitted, the phase information is unchanged and the activation is effectively a masking function as in Section \ref{subsec:cvmask}. 
Additionally, a generalized Type-B activation function is included allowing the user to adopt any set of $G_{||}(\cdot)$ and $G_\angle(\cdot)$ desired. 

\begin{table}[ht]
    \centering
    \begin{tabular}{c|c|c|c}
        \href{https://complextorch.readthedocs.io/en/latest/index.html}{\texttt{complextorch}} Activation Layer & $G_{||}(|\mathbf{z}|)$ & $G_\angle(\angle\mathbf{z})$ & Reference  \\
        \hline
        \texttt{CVPolarTanh} & $\tanh(|\mathbf{z}|)$ & - & Eq. (8) \cite{hirose2012generalization} \\
        \hline
        \texttt{CVPolarSquash} & $\frac{|\mathbf{z}|^2}{(1 + |\mathbf{z}|^2)}$ & - & Section III-C \cite{hirose2012generalization} \\
        \hline
        \texttt{CVPolarLog} & $\ln(|\mathbf{z}| + 1)$ & - & Section III-C \cite{hayakawa2018applying} \\
        \hline
        \texttt{modReLU} & $\texttt{ReLU}(|\mathbf{z}| + b)$ & - & Eq. (8) \cite{arjovsky2016modrelu} \\
        \hline
    \end{tabular}
    \caption{Type-B activation functions.}
    \label{tab:typeB}
\end{table}

\subsubsection{Fully Complex Activation Layers}

Fully complex activation layers employ activation functions specifically designed for complex-valued data and hence do not have a general form. 
Table \ref{tab:fully_complex_act} details the fully-complex activation functions included in \texttt{complextorch}. 

\begin{table}[ht]
    \centering
    \begin{tabular}{c|c|c}
        \href{https://complextorch.readthedocs.io/en/latest/index.html}{\texttt{complextorch}} Activation Layer & $G(\mathbf{z})$ & Reference \\
        \hline
        \texttt{CVSigmoid} & $\frac{1}{1 + \exp{(\mathbf{z})}}$ & Eq. (71) \cite{nitta2017hyperbolic} \\
        \hline
        \texttt{zReLU} & $\begin{cases} \mathbf{z} \quad \text{if} \quad \angle\mathbf{z} \in [0, \pi/2] \\ 0 \quad \text{else} \end{cases}$ & Section 4.2.1 \cite{guberman2016zrelu} \\
        \hline
        \texttt{CVCardiod} & $\frac{1}{2} (1 + \text{cos}(\angle\mathbf{z})) \odot \mathbf{z}$ & Eq. (3) \cite{virtue2017cardioid} \\
        \hline
        \texttt{CVSigLog} & $\frac{\mathbf{z}}{(c + 1/r * |\mathbf{z}|)}$ & Eq. (20) \cite{georgiou1992complex} \\
        \hline
    \end{tabular}
    \caption{Fully-complex activation functions.}
    \label{tab:fully_complex_act}
\end{table}

\subsubsection{ReLU-Based Complex-Valued Activation Layers}

The ReLU is the most popular activation function in modern deep learning, and it has garnered significant attention in its extension to the complex domain. 
Most take a similar form to Type-A activation functions as in Section \ref{subsubsec:typeA}, operating on the real and imaginary parts of the input signal. 
However, some functions, like the Type-B \texttt{modReLU} and fully-complex Guberman ReLU (\texttt{zReLU}) apply ReLU-like operations. 
Some efforts have been made to develop insights into how ReLU-based complex-valued activation functions ``activate'' across different regions and quadrants of the complex plane \cite{jing2022enhanced,guberman2016zrelu}. 
Table \ref{tab:complex_relu} details the fully-complex activation functions included in \texttt{complextorch}, where $\mathbf{z} = \mathbf{x} + j\mathbf{y}$. 

\begin{table}[ht]
    \centering
    \begin{tabular}{c|c|c}
        \href{https://complextorch.readthedocs.io/en/latest/index.html}{\texttt{complextorch}} Activation Layer & $G(\mathbf{z})$ & Reference \\
        \hline
        \texttt{CVSplitReLU} & $\texttt{ReLU}(\mathbf{x}) + j \texttt{ReLU}(\mathbf{y})$ & Eq. (5) \cite{gao2018enhanced} \\
        \hline
        \texttt{CReLU} & $\texttt{ReLU}(\mathbf{x}) + j \texttt{ReLU}(\mathbf{y})$ & Eq. (5) \cite{gao2018enhanced} \\
        \hline
        \texttt{CPReLU} & $\texttt{PReLU}(\mathbf{x}) + j \texttt{PReLU}(\mathbf{y})$ & Eq. (2) \cite{jing2022enhanced} \\
        \hline
    \end{tabular}
    \caption{ReLU-based activation functions.}
    \label{tab:complex_relu}
\end{table}

\subsection{Complex-Valued Loss Functions}
\label{subsec:cvloss}

In this section, we overview some common complex-valued loss functions. 
Whereas we emphasize regression loss, complex-valued classification and other loss functions are further explored in \cite{barrachina2023theory,lee2022complex}. 
Similar to activation functions, two general types of loss functions have similar forms to Type-A and Type-B activations, operating on the real and imaginary or magnitude and phase, respectively. 

\subsubsection{Split Loss Functions}

Split loss functions apply two real-valued loss functions to the real and imaginary parts of the estimated ($\mathbf{x})$) and ground truth ($\mathbf{y})$) labels as
\begin{equation}
    \mathcal{L}(\mathbf{x}, \mathbf{y}) = \mathcal{L}_\mathbb{R}(\mathbf{x}_\mathbb{R}, \mathbf{y}_\mathbb{R}) + \mathcal{L}_\mathbb{I}(\mathbf{x}_\mathbb{I}, \mathbf{y}_\mathbb{I}),
\end{equation}
where the total loss is computed as the sum of the real and imaginary losses. 
Generally $\mathcal{L}_\mathbb{R}(\cdot, \cdot) = \mathcal{L}_\mathbb{I}(\cdot, \cdot)$; however, we include an generalized split loss function allowing the user to specify any combination of $\mathcal{L}_\mathbb{R}(\cdot, \cdot)$ and $\mathcal{L}_\mathbb{I}(\cdot, \cdot)$. 
Table \ref{tab:split_loss} details the split loss functions included in \texttt{complextorch}. 

\begin{table}[ht]
    \centering
    \begin{tabular}{c|c}
        \href{https://complextorch.readthedocs.io/en/latest/index.html}{\texttt{complextorch}} Loss Function & $\mathcal{L}_\mathbb{R}(\cdot, \cdot) = \mathcal{L}_\mathbb{I}(\cdot, \cdot)$ \\
        \hline
        \texttt{SplitL1} & $\texttt{L1}(\cdot,\cdot)$ \\
        \hline
        \texttt{SplitMSE} & $\texttt{MSE}(\cdot,\cdot)$ \\
        \hline
        \texttt{SplitSSIM} & $\texttt{SSIM}(\cdot,\cdot)$ \\
        \hline
    \end{tabular}
    \caption{Split Activation Functions.}
    \label{tab:split_loss}
\end{table}

\subsubsection{Polar Loss Functions}

Similarly, polar loss functions apply two real-valued loss functions to the magnitude and phase of the estimated and ground truth labels as
\begin{equation}
    G(\mathbf{x}, \mathbf{y}) = w_{||} G_{||}(|\mathbf{x}|, |\mathbf{y}|) + w_\angle G_\angle(\angle\mathbf{x}, \angle\mathbf{y}),
\end{equation}
where $w_{||}$ and $w_\angle$ are scalar weights applied based on \textit{a priori} understanding of the problem to scale the magnitude and phase losses, particularly as the phase loss will always be less than $2 \pi$ by definition. 
Tuning the loss weights may improve modeling performanceas 
To the author's understanding, at the time of this paper, there have been no efforts to apply polar loss functions. 
However, they may, in conjunction with split loss functions, improve modeling performance by imposing additional loss on the phase-accuracy of the algorithm. 

\subsubsection{Other Complex-Valued Loss Functions} 

Several complex-valued loss functions are implemented and detailed in Table \ref{tab:complex_loss}
\begin{table}[ht]
    \centering
    \begin{tabular}{c|c|c}
        \href{https://complextorch.readthedocs.io/en/latest/index.html}{\texttt{complextorch}} Activation Layer & $G\mathcal{L}(\mathbf{x}, \mathbf{y})$ & Reference \\
        \hline
        \texttt{CVQuadError} & $\frac{1}{2}\text{sum}(|\mathbf{x} - \mathbf{y}|^2)$ & Eq. (11) \cite{haensch2010complex} \\
        \hline
        \texttt{CVFourthPowError} & $\frac{1}{2}\text{sum}(|\mathbf{x} - \mathbf{y}|^4)$ & Eq. (12) \cite{haensch2010complex} \\
        \hline
        \texttt{CVCauchyError} & $\frac{1}{2}\text{sum}( c^2 / 2 \ln(1 + |\mathbf{x} - \mathbf{y}|^2/c^2) )$ & Eq. (13) \cite{haensch2010complex} \\
        \hline
        \texttt{CVLogCoshError} & $\text{sum}(\ln(\cosh(|\mathbf{x} - \mathbf{y}|^2))$ & Eq. (14) \cite{haensch2010complex} \\
        \hline
        \texttt{CVLogError} & $\text{sum}(|\ln(\mathbf{x}) - \ln(\mathbf{y})|^2)$ & Eq. (10) \cite{bassey2021survey} \\
        \hline
    \end{tabular}
    \caption{Other complex-valued loss functions.}
    \label{tab:complex_loss}
\end{table}
Additionally, \texttt{PerpLossSSIM} (Eq. (5), Fig. 1 \cite{terpstra2022loss}) is implemented.
Its mathematical formulation is omitted here, but can be found in \cite{terpstra2022loss}. 

\subsection{Complex-Valued Manifold-Based Layers}
\label{subsec:manifold}

In \cite{chakraborty2019sur_real,chakraborty2022SurReal} a complex-valued convolution operator offering similar equivariance properties to the spatial equivariance of the traditional real-valued convolution operator is introduced.
By approaching the complex domain as a Riemannian homogeneous space consisting of the product of planar rotation and non-zero scaling, they define a convolution operator equivariant to phase shift and amplitude scaling.
Although their paper shows promising results in reducing the number of parameters of a complex-valued network for several problems, their work has not gained mainstream support. 
However, some initial work has shown significant promise in reducing model sizes and improving modeling capacity for smaller models \cite{brown2021charrnets,liu2021mmff,scarnati2021complex}.
Incorporating manifold-based complex-valued deep learning is a promising research area for future efforts. 
For full derivation and alternative implementations, please refer to \cite{chakraborty2019sur_real,chakraborty2020C_SURE,chakraborty2022SurReal}. 

As the authors mention in the final bullet point in Section IV-A1,
\begin{displayquote}
    ``If $d$ is the manifold distance in (2) for the Euclidean space that is also Riemannian, then wFM has exactly the weighted average as its closed-form solution. That is, our wFM convolution on the Euclidean manifold is reduced to the standard convolution, although with the additional convexity constraint on the weights.''
\end{displayquote}
Hence, the implementation closely follows the conventional convolution operator with the exception of the weight normalization.
We would like to not that the weight normalization, although consistent with the authors' implementation, lacks adequate explanation from the literature and could be improved for further clarity. 

\texttt{complextorch} contains implementations for 1-D and 2-D versions of the proposed wFM-based convolution operated introduced in \cite{chakraborty2019sur_real,chakraborty2022SurReal}, dubbed \texttt{wFMConv1d} and \texttt{wFMConv2d}, respectively. 

\subsection{Complex-Valued Pooling Layers}
\label{subsec:cvpool}

Complex-valued average pooling can be computed similarly to real-valued pooling where the average can be computed over the real and imaginary parts of the signal separately as
\begin{equation}
    \texttt{CVAdaptiveAvgPoolingNd}(\mathbf{z}) = \texttt{AdaptiveAvgPoolingNd}(\mathbf{x}) + j \texttt{AdaptiveAvgPoolingNd}(\mathbf{y}),
\end{equation}
where $\mathbf{z} = \mathbf{x} + j\mathbf{y}$. 
We include implementations for \texttt{CVAdaptiveAvgPooling1d}, \texttt{CVAdaptiveAvgPooling2d}, and \texttt{CVAdaptiveAvgPooling3d}. 

\subsection{Complex-Valued Dropout Layers}
\label{subsec:cvdropout}

Similarly, complex-valued dropout can be computed similarly to real-valued dropout where dropout can be computed over the real and imaginary parts of the signal separately as
\begin{equation}
    \texttt{CVDropout}(\mathbf{z}) = \texttt{Dropout}(\mathbf{x}) + j \texttt{Dropout}(\mathbf{y}),
\end{equation}
where $\mathbf{z} = \mathbf{x} + j\mathbf{y}$. 

\section{Conclusion}
\label{sec:conclusion}

In this paper, we introduced a PyTorch wrapper for complex-valued neural network modeling. 
The proposed framework enables rapid development of deep learning models for signal processing and signal understanding tasks relying on complex-valued data. 
We detailed the implementation of deep learning layers spanning convolution, linear, activation, attention, and loss functions. 

%\listofalgorithms
% References should be produced using the bibtex program from suitable
% BiBTeX files (here: strings, refs, manuals). The IEEEbib.bst bibliography
% style file from IEEE produces unsorted bibliography list.
% -------------------------------------------------------------------------
\bibliographystyle{IEEEbib}
\bibliography{mega_bib}
% PSNR_{together} & $\textbf{36.50}$ & $26.23$ & $21.14$ & $21.15$\\ 
% RMSE_{together} & $\textbf{0.224}$ & $1.47$ & $1.324$ & $1.32$\\

\end{document}